# A bone suppression model ensemble to improve COVID-19 detection in chest X-rays

**Sivaramakrishnan Rajaraman[1*], Gregg Cohen[2], Lillian Spear[2], Les folio[2], and Sameer Antani[1]**

[1] National Library of Medicine, National Institutes of Health, Maryland, USA

[2] Clinical Center, Department of Radiology and Imaging Sciences, National Institutes of Health, Maryland, USA

* Correspondence to:

Sivaramakrishnan Rajaraman, PhD., TEL: +1-301-827-2383 ; E-mail: sivaramakrishnan.rajaraman@nih.gov

## Abstract

Chest X-ray (CXR) is a widely performed radiology examination that helps to detect abnormalities in the tissues and organs in the thoracic cavity. Detecting pulmonary abnormalities like COVID-19 may become difficult due to that they are obscured by the presence of bony structures like the ribs and the clavicles, thereby resulting in screening/diagnostic misinterpretations. Automated bone suppression methods would help suppress these bony structures and increase soft tissue visibility. In this study, we propose to build an ensemble of convolutional neural network models to suppress bones in frontal CXRs, improve classification performance, and reduce interpretation errors related to COVID-19 detection. We train and evaluate variants of U-Nets, feature pyramid networks, and other proposed custom models using a private collection of CXR images and their bone-suppressed counterparts. The ensemble is constructed by (i) measuring the multi-scale structural similarity index (MS-SSIM) score between the sub-blocks of the bone-suppressed image predicted by each of the top-3 performing bone-suppression models and the corresponding sub-blocks of its respective ground truth soft-tissue image, and (ii) performing a majority voting of the MS-SSIM score computed in each sub-block to identify the sub-block with the maximum MS-SSIM score and use it in constructing the final bone-suppressed image. We empirically determine the sub-block size that delivers superior bone suppression performance. It is observed that the bone suppression model ensemble outperformed the individual models in terms of PSNR (36.7977±1.6207), MS-SSIM (0.9848±0.0073), and other metrics. The best-performing bone-suppression model is used to suppress the bones in publicly available CXR collections. A CXR modality-specific classification model is retrained and evaluated on the non-bone-suppressed and bone-suppressed images to classify them as showing normal lungs or other COVID-19-like manifestations. We observed that the bone-suppressed model training significantly outperformed (MCC: 0.9645, 95% confidence interval (0.9510, 0.9780)) ($p < 0.05$) the model trained on non-bone-suppressed images (MCC: 0.7961, 95% confidence interval (0.7667, 0.8255)) toward detecting COVID-19 manifestations, signifying that bone suppression improved the model sensitivity toward COVID-19 classification.

## I.   Introduction

Chest X-ray (CXR) is a commonly performed radiological examination to reveal the conditions of the lungs, heart, ribs, blood vessels, and monitor post-operation recovery. However, accurate interpretation of pulmonary abnormalities like COVID-19 and others is particularly challenging because their visibility may be obstructed by the presence of bony structures like ribs and clavicles. This reduced visibility may lead to an erroneous interpretation by an expert or an automated artificial intelligence (AI) algorithm, severely impacting clinical decision-making. Literature studies reveal that the presence of ribs and clavicles in CXR images led to missed lung cancer interpretations (Shah et al., 2003). Advanced radiology methods like dual-energy subtraction (DES) chest radiography are used to produce bone-selective and soft tissue-selective images (Manji et al., 2016). However, compared to traditional CXRs, DES has several limitations (Kuhlman et al., 2006): (i) DES radiography exposes the subjects to a slightly higher radiation dosage compared to traditional CXR imaging; (ii) DES radiographic imaging can only be performed in the posterior-anterior view; (iii) Cost of DES radiography is higher compared to conventional CXR imaging; (iv) Lack of portability could adversely impact its use in low and middle resource regions (LMRR), and (v) DES radiography is recommended only for patients above 16 years of age. It is, therefore, indispensable to propose automated bone suppression methods that suppress the bony structures present in traditional CXRs, enhance soft tissue visibility, and improve the detection of pulmonary manifestations.

A study of the literature reveals several works published on suppressing bones in CXRs. These studies involve using (i) commercial software, (ii) conventional machine learning methods using hand-crafted feature descriptors, or (iii) state-of-the-art (SOTA) deep learning (DL) models to initially generate bone-selective images and further subtract them from the original CXR to increase soft-tissue visibility. In (Li et al., 2011b), the authors used commercial software to suppress bones and improve performance in detecting lung nodules. It was observed that the performance of the experts significantly improved ($p < 0.05$) by obtaining an area under the receiver-operating-characteristic curve (AUROC) of 0.863 using the bone-suppressed CXRs, compared to an AUC of 0.82 using non-bone-suppressed CXRs. Another study used commercial bone suppression software to suppress bones in CXRs and investigated for a performance improvement in TB detection (Kodama et al., 2018). It was observed that the average AUC of experts improved from 0.882 to 0.933 using bone-suppressed images. A convolutional neural network (CNN)-based model was used in (Matsubara et al., 2019) to generate a bone-selective image. The generated image is subtracted from the original CXR to increase soft-tissue visibility. The authors achieved 89.2% bone suppression in this regard. A cascade of CNNs was used in (Yang et al., 2017) to create bone-selective images at multiple scales. The generated images were fused to form the final bone-selective image that was subtracted from the original CXR to generate a "bone-free" image. In another study (Suzuki et al., 2006), an artificial neural network was used to generate a bone-selective image that was subtracted from the original CXR to increase the visibility of soft tissues. A method based on independent component analysis was proposed in (Nguyen and Dang, 2015) to suppress bones and increase lung nodule visibility. Other studies (Freedman et al., 2011; Li et al., 2012, 2011a; Oda et al., 2009) adopted bone suppression methods to improve the performance toward detecting lung nodules and other pulmonary manifestations. These studies in general, propose multiple steps to generate bone-selective images and further subtract them from the original CXRs to increase soft-tissue visibility. A limitation of this approach is that an inaccurate generation of bone-selective images would lead to introducing noise, reducing the visibility of soft tissues, increasing interpretation errors, and adversely impacting decision-making. Except for (Rajaraman et al., 2021), the literature is limited considering proposing an automated method that generates a soft-tissue image directly from the original CXR image, alleviating the need for intermediate bone-selective generation and subtraction methods.

Though CNN models demonstrate SOTA performance in natural and medical vision recognition tasks, they are often found to suffer from bias and variance issues that could adversely affect their interpretation. These issues could be tackled through ensemble learning that optimally combines the predictions of several models to improve prediction performance compared to the individual constituent models and reduce prediction spread or dispersion (Dietterich, 2000). Ensemble learning is widely used in medical computer vision tasks such as segmentation, object detection, and classification (Rajaraman et al., 2020). To the best of our knowledge, we observed that no literature exists considering evaluating the performance of DL models ensembles toward bone suppression in CXRs.

In this study, we propose the benefits offered through (i) building an ensemble of DL models to suppress bones in frontal CXRs and improve classification and interpretation performance related to COVID-19 detection. We train several SOTA architectures such as U-Nets (Ronneberger et al., 2015) and Feature Pyramid Networks (FPNs) (Xie et al., 2018), using several ImageNet classifier backbones, and also propose custom models toward bone suppression. A bone suppression model ensemble is constructed by (i) measuring the multi-scale structural similarity index (MS-SSIM) score between the sub-blocks of the bone-suppressed image predicted by each of the top-3 performing bone-suppression models and the corresponding sub-blocks of its respective ground truth soft-tissue image, and (ii) performing a majority voting of the MS-SSIM score computed in each sub-block to identify the sub-block with the maximum MS-SSIM score and use it in constructing the final bone-suppressed image. We empirically determine the sub-block size that delivers superior bone suppression performance. The performances of individual models and the model ensemble are evaluated using several performance metrics such as average peak signal-to-noise (PSNR) ratio, structural similarity index (SSIM), MS-SSIM, correlation, intersection, chi-square, and Bhattacharya distances. The best performing bone suppression model is used to suppress bones in publicly available CXR collections. The best-performing bone suppression model is truncated and appended with classification layers to transfer CXR modality-specific knowledge and improve performance in the task of classifying CXRs as showing normal lungs or other COVID-19-related manifestations. The performances of the classification model trained on non-bone-suppressed CXRs and bone-suppressed CXRs are compared through several performance metrics such as accuracy, AUROC, precision, recall, the area under the precision-recall curve (AUPRC), F-score, and MCC. Additionally, we used our in-house class-selective relevance map (CRM) localization algorithm (Incheol Kim et al., 2019) to interpret model predictions. Fig. 1 shows the graphical abstract of our proposed approach.

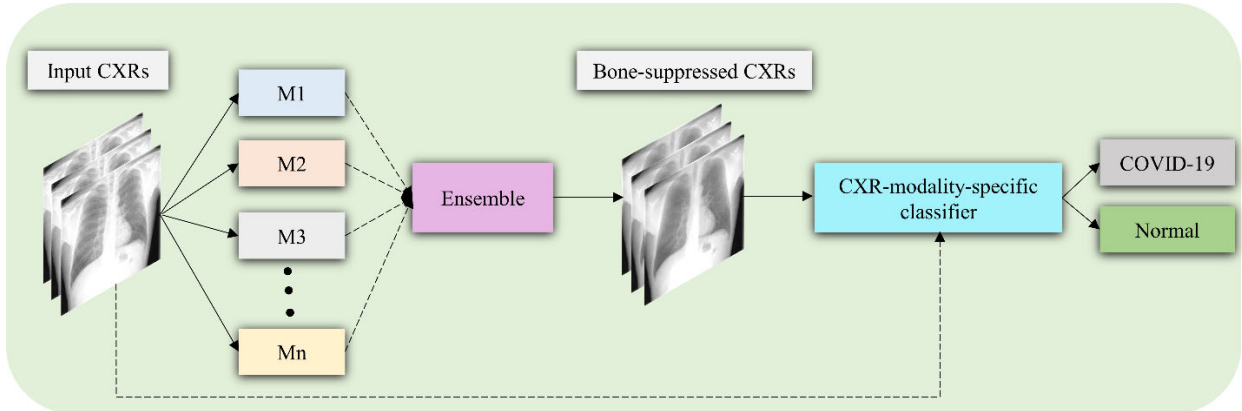

**Fig. 1. Graphical abstract of the proposal.** The input CXRs are fed into the proposed bone suppression models (M1, M2, …, M$n$, $n$=1, 2, …, 14). A model ensemble is constructed by combining the predictions of the top-3 performing bone suppression models using a majority voting approach. A classification model is trained on the non-bone-suppressed and bone-suppressed images to classify them into COVID-19 or normal categories.

The contributions of this retrospective study are highlighted as follows:

(i) We are releasing a collection of 127 DES images (original CXRs, bone-selective images, and soft-tissue images), post publication, to promote global research. To the best of our knowledge, this would be the first publicly available collection of DES images that would benefit the research community to help train bone-suppression models.

(ii) To the best of our knowledge, this would be the first study to evaluate the performance of a model ensemble toward suppressing bones in CXRs.

(iii) We performed extensive empirical evaluations, statistical significance analyses, qualitative and quantitative evaluations to demonstrate the efficacy of the proposed approach.

(iv) The individual constituent models and their ensemble proposed in this study are not restricted to the task of CXR bone suppression but can be potentially applied to other image-denoising applications.

The rest of the study is organized as follows: Section II discusses the datasets and methods used, Section III interprets the results, and Section IV discusses and concludes this study.

## II. Materials and methods

**Datasets**

This retrospective study uses the following datasets:

(i) COVID-19 CXR collection: A total of 3016 CXR images showing COVID-related manifestations are pooled from several publicly available datasets, GitHub repositories, and other sources. A majority of these CXRs are pooled from the publicly available BIMCV-COVID19+ CXR data collection that contains 2473 CXRs showing COVID-19-related manifestations (Vayá et al., 2020). A set of 183 CXR images showing COVID-19-like manifestations are collected from a GitHub repository hosted by the Institute for Diagnostic and Interventional Radiology, Hannover Medical School, Hannover, Germany (Institute for Diagnostic and Interventional Radiology, 2020). The CXR images are accompanied by other metadata such as admission status and patient demographics. The authors of (Rajaraman et al., 2020) collected 226 CXRs showing COVID-19-related manifestations, from a public GitHub repository hosted by the authors of (Cohen et al., 2020). The CXR collection is accompanied by other metadata including sex, age, finding, and intubation status. The authors of (Rajaraman et al., 2020) used a collection of 134 CXRs acquired from SARS-CoV-2 PCR+ patients from a hospital in Spain and posted by a radiologist in a public Twitter thread (Imaging, 2020). The ground truth COVID-19 disease-specific ROI annotations, set by the verification from two expert radiologists, for a subset of this collection [n = 36] are used by the authors of [16] in interpreting model performance.

(ii) RSNA CXR dataset: An equal number of 3016 CXR images showing no abnormalities are pooled from the publicly available RSNA CXR dataset, released toward the RSNA pneumonia detection challenge hosted by Kaggle (Shih et al., 2019). The collection, however, includes a total of 26,684 CXR images, of which, 8851 CXRs showed no abnormalities, 6012 CXRs showed pneumonia-related lung opacities, and 11, 821 CXRs showed other pulmonary abnormalities.

(iii) NIH-CC-DES-Set 2: A set of 100 DES CXRs are acquired using the GE Discovery XR656 digital radiography system, by the NIH Clinical Center, from patients as a part of routine clinical care. The DES images are captured at 120kVp and 50kVp to respectively capture the soft-tissue and bony-selective images. The collection contains DES images of 54 females and 46 males, the average age and standard deviation of the males and females are 48.9 +/- 14.5 and 45.4+/- 13.6, respectively. All DES images are labeled as normal from experts' interpretations. The dataset is augmented and further used to train the bone suppression models.

(iv) NIH-CC-DES-Set 1: The authors of (Rajaraman et al., 2021) used a collection of 27 DES CXR images that were acquired using the GE Discovery XR656 digital radiography system as a part of routine clinical care. The DES images were taken at 120 and 50 Kilovoltage-peak (kVp) to respectively capture the soft-tissue images and bony-selective images. We used this as the test set to evaluate the performance of the proposed bone suppression models. The total number of CXRs pooled from different sources is given in Table 1.

**Table 1. Dataset sources.**

| Source | Number of CXR images | |
|---|---|---|
| | COVID-19 | Normal |
| BIMCV-COVID19+ CXR | 2473 | NA |
| Hannover Medical School, Hannover | 183 | NA |
| Cohen et al. | 226 | NA |
| Twitter COVID-19 CXR | 134 | NA |
| RSNA CXR | 3016 | 3016 |
| NIH-CC-DES-Set 1 | NA | 27 |
| NIH-CC-DES-Set 2 | NA | 100 |

## Bone suppression models

The set of 100 grayscale DES CXR images (i.e., the original CXRs and soft tissue counterparts) from the NIH-CC-DES-Set 2 dataset are augmented using affine transformations such as rotations (-10 to 10 degrees), horizontal and vertical shifting (-5 to 5 pixels), horizontal mirroring, zooming, median filtering, Gaussian blurring, and unsharp masking, resulting in 1000 DES CXRs. The augmented images are further resized to 256×256 dimensions to reduce computational complexity. The contrast of the images is enhanced by saturating the bottom and top 1% of all image pixel values. The pixel values are then normalized. We propose the following model architectures for the task of bone suppression in CXRs: (i) Autoencoder-BS (Bone Suppression); (ii) ResNet-BS; (iii) U-EB0-BS; (iv) U-Res18-BS; (v) U-SE-Res18-BS; (vi) U-D121-BS; (vii) U-IV3-BS; (viii) U-MobileV2-BS; (ix) FPN-EB0-BS; (x) FPN-Res18-BS; (xi) FPN-SE-Res18-BS; (xii) FPN-D121-BS, (xiii) FPN-IV3-BS; (xiv) FPN-MobileV2-BS. The model architectures are discussed in subsequent sections.

**Autoencoder with separable convolutions (Autoencoder-BS)**

The Autoencoder-BS model is a convolutional denoising autoencoder with symmetrical encoder and decoder layers. Fig. 2 illustrates the architecture of the proposed Autoencoder-BS model.

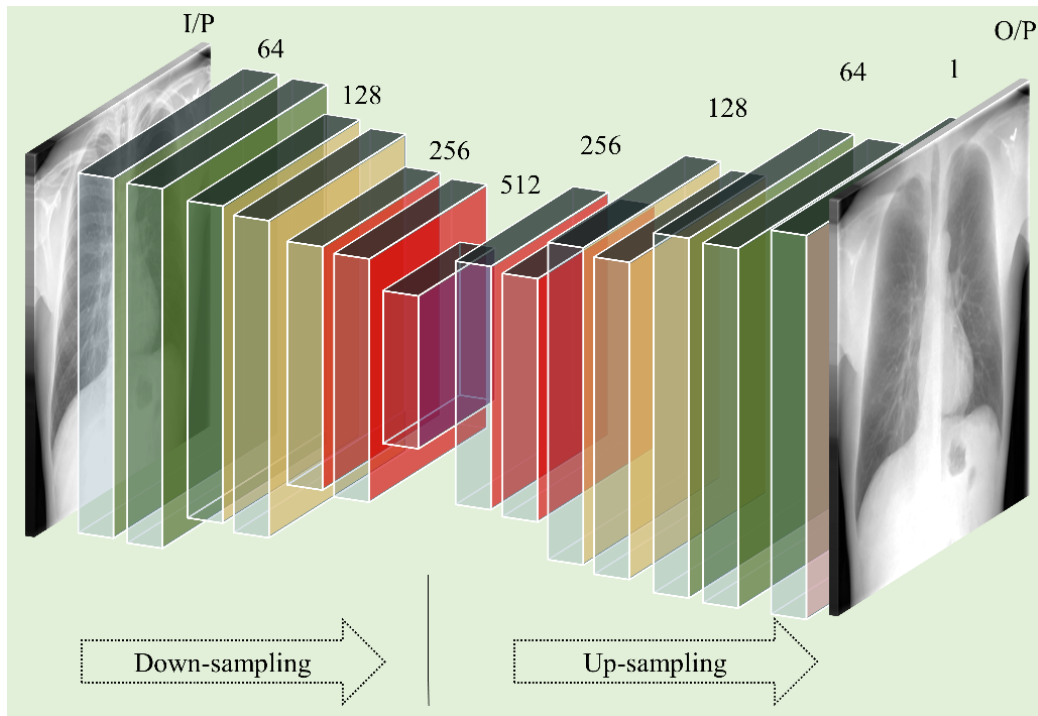

**Fig. 2. The architecture of the Autoencoder-BS model.** The input to the model is a grayscale CXR image. The model has a symmetrical separable convolutional encoder and decoder architecture.

The encoder consists of four separable convolutional blocks. Each convolutional block except for the last block contains two separable convolutional layers. In a separable convolution layer, a depth-wise spatial convolution followed by a pointwise convolution is performed on the input channels to mix the resulting outputs. Such an approach is reported to reduce computational complexity compared to traditional convolution operations, thereby facilitating faster convergence and real-time deployment (Chollet, 2016). The number of filters in the separable convolutional blocks of the encoder are 64, 128, 256, and 512, respectively. Except for the last block, a max-pooling layer is used after each separable convolutional block to calculate the maximum value for individual patches of the feature map. Upsampling layers are used correspondingly in the symmetric decoder blocks to preserve the spatial resolution of the input.

**ResNet-based model with residual scaling (ResNet-BS)**

The architecture of the proposed ResNet-BS model is shown in Fig. 3. The first and last convolutional layer contains 128 filters of dimension 3×3. We used residual blocks with shortcuts to skip over layers. This approach helps to overcome convergence issues due to vanishing gradients in deeper models. Skipping layers helps to reuse the activations of the earlier layers until weight updates in the succeeding layers. Each residual block consists of two convolutional layers with 3×3 filters and 128 feature maps.

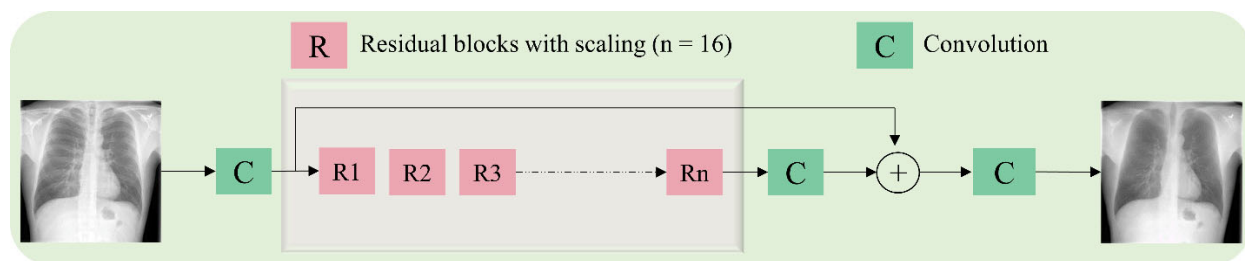

**Fig. 3. The architecture of the ResNet-BS model**. The input is a grayscale CXR image. The block *R* denotes the modified residual block. The proposed model has 16 residual blocks, each having two convolutional layers with 3×3 filters and 128 feature maps. The deepest convolutional layer with sigmoidal activation predicts the grayscale bone-suppressed image.

Inspired by (Lim et al., 2017), we used a modified residual block in which (i) the batch normalization layers are removed for they are mentioned to adversely affect the range flexibility through the normalization process, and (ii) activations are not used outside the residual blocks and in the final layer. The network consists of 16 residual blocks with an identical layout. We used zero paddings to preserve the spatial dimensions of the input image. The residuals after the deepest convolutional layer in each residual block are scaled at an empirically determined scaling factor (0.1) before adding them back to the convolutional path. This scaling approach stabilizes training in deeper models with high computational complexity (Lim et al., 2017). The deepest convolutional layer with the sigmoidal activation function predicts a grayscale bone-suppressed image.

**U-Net and FPN-based models**

The U-Net models are widely used in image segmentation tasks (Ronneberger et al., 2015). The U-Net is composed of an encoder and decoder. The encoder or the contracting path extracts image features at multiple scales and the decoder or the expanding path semantically projects the features learned by the encoder onto the pixel space. The feature pyramid networks (FPN) are widely used as feature extractors to help object detection (Xie et al., 2018). Fig. 4 shows the general architecture of the U-Net and FPN models.

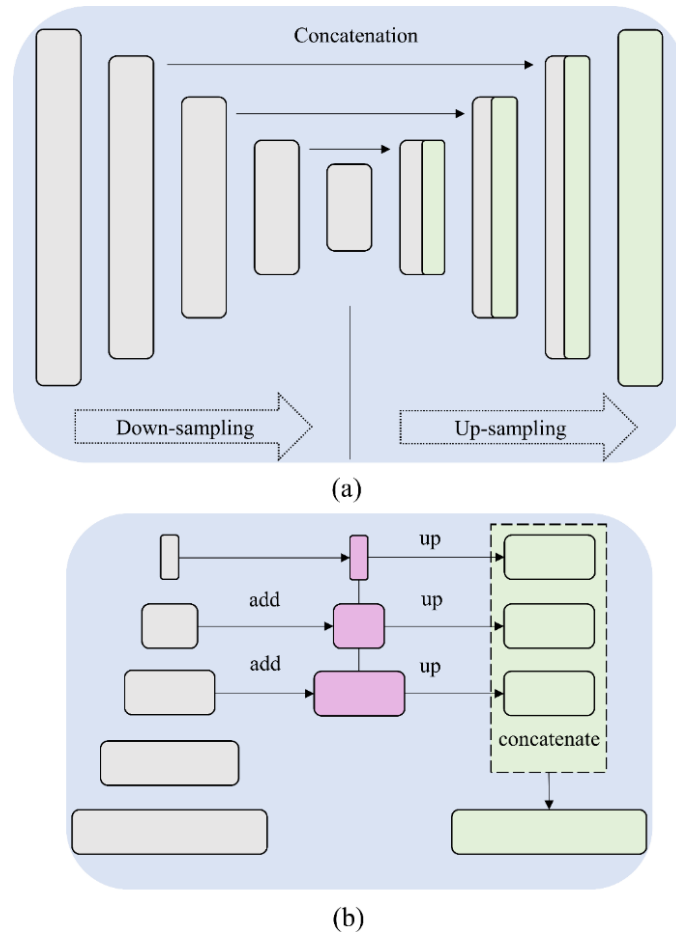

**Fig. 4. The general architecture of the (a) U-Net and (b) FPN model.** The input is a three-channel CXR image. The following encoder-backbones are used in this study (EfficientNet-B0, ResNet-18, SE-ResNet-18, DenseNet-121, Inception-V3, MobileNet-V2, trained on ImageNet dataset)

The FPN network is composed of bottom-up and top-down pathways. The bottom-up pathway constitutes the encoder backbone that extracts image features at multiple scales (scaling step is 2). A convolutional layer with a 1×1 filter is used to reduce the feature dimensions of the deepest convolutional layer in the bottom-up pathway to 256. This constitutes the first layer of the top-down pathway. Going deeper, the preceding layer is up-sampled by a factor of 2 using the nearest neighbor up-sampling method. A 1×1 convolutional filter is applied to the corresponding feature maps in the bottom-up pathway and is added element-wise. A 3×3 convolution is then applied to all the merged layers to reduce aliasing effects. This helps to generate high-resolution features at each scale.

The grayscale CXR is duplicated in three channels and fed into the U-Net and FPN models. This is because we use ImageNet-pretrained models, trained on RGB images, as the encoder backbones. We experimented with several encoder backbones for the U-Net and FPN models (Pavel Yakubovskiy, 2020) toward the task of bone suppression in CXRs. These backbones include (i) EfficientNet-B0 (Tan and Le, 2019), (ii) ResNet-18 (He et al., 2016), (iii) SE-ResNet-18 (Hu et al., 2018), (iv) DenseNet-121 (Huang et al., 2017), (v) Inception-V3 (Szegedy et al., 2016), and (vi) MobileNet-V2 (Sandler et al., 2018). We are motivated by the fact that these ImageNet-pretrained models have demonstrated superior performance in medical visual recognition tasks (Rajaraman et al., 2020). The final layer of the U-Net and FPN models consists of a convolutional layer with Sigmoidal activation to predict grayscale bone-suppressed CXRs.

The proposed bone-suppression models are trained on the augmented NIH-CC-DES-Set 2 dataset and tested with the NIH-CC-DES-Set 1 dataset. We allocated 10% of the training data for validation using a fixed seed. An Ubuntu Linux system with NVIDIA GeForce GTX 1080 graphics card and Keras framework with Tensorflow backend is used for model training and evaluation. We compiled the models using an Adam optimizer with an initial learning rate of 1e-3 and monitored the following validation performance metrics: (i) loss, (ii) PSNR, (iii) SSIM, and (iv) MS-SSIM. We propose a mixed-loss function that benefits from the combination of mean absolute error (MAE) and MS-SSIM losses, given by,

$$Mixed\ loss = \Omega.MS-SSIM + (1-\Omega).MAE$$

We empirically set the value of $\Omega$ to 0.84. The MS-SSIM metric is given higher weightage since we prefer the bone suppressed image to have the least structural alterations and high similarity to the ground truth. The MAE metric is given less weightage comparatively because this loss component focusses on the overall contrast and luminance in the image and these factors are expected to vary while suppressing the bones (white pixels). We used callbacks to store model checkpoints and reduced the learning rate when the validation performance did not improve. We used early stopping with the patience of 10 epochs and checked the validation performance. The best-performing models (with the least validation loss) are further used to predict bone-suppressed CXR images using the test set.

**Bone suppression model ensemble**

The bone suppression model ensemble is constructed using the top-3 performing models that demonstrate markedly improved performance in terms of the MS-SSIM metric using the NIH-CC-DES-Set 1 test set. Each of the top-3 performing models predicts a bone-suppressed image for an input CXR. The predicted image by the individual models is divided into sub-blocks of M×M dimensions. The optimal value of M [4, 8, 16, 32, 64, 128, 256] is determined through extensive empirical evaluations. For a given sub-block size and in each sub-block, the following steps are performed: (i) we measured the MS-SSIM score between the sub-block of the bone-suppressed image predicted by each of the top-3 performing models and the corresponding sub-block of its respective ground truth soft-tissue image; (ii) we performed a majority voting for the MS-SSIM score to find that image sub-block with the maximum MS-SSIM score and use it in constructing the final bone-suppressed image. The algorithm below discusses these steps. Fig. 5 illustrates the architecture of the proposed BSE model.

**Algorithm**

**Input:** Ground-truth bone-suppressed image $K$ of $256 \times 256$ resolution

Bone-suppressed Images $I = (I_{M1}, I_{M2}, I_{M3})$ of $256 \times 256$ resolution from $M = [M_1, M_2, M_3]$ ; $M_1, M_2, M_3$ are the top-3 performing bone-suppression models

Image sub-block sizes $B = [4, 8, 16, 32, 64, 128, 256]$

**Output:** Final Bone-suppressed image $J$

*for* each sub-block size $B$

    *for* each set of bone-suppressed Images $I$ generated by $M_1, M_2, M_3$

        *for* each sub-block in $K$ and $I_{M1}, I_{M2}, I_{M3}$

            compute **MS-SSIM** between $K$ and $I_{M1}$, $K$ and $I_{M2}$, $K$ and $I_{M3}$

perform *Majority Voting* = *Max(MS-SSIM(K* and *I$_{M1}$), MS-SSIM(K* and *I$_{M2}$), MS-SSIM(K* and *I$_{M3}$))*

choose the sub-block with the maximum *MS-SSIM* value and put it in its respective position in the final bone-suppressed image *J*

*end for*

*end for*

*end for*

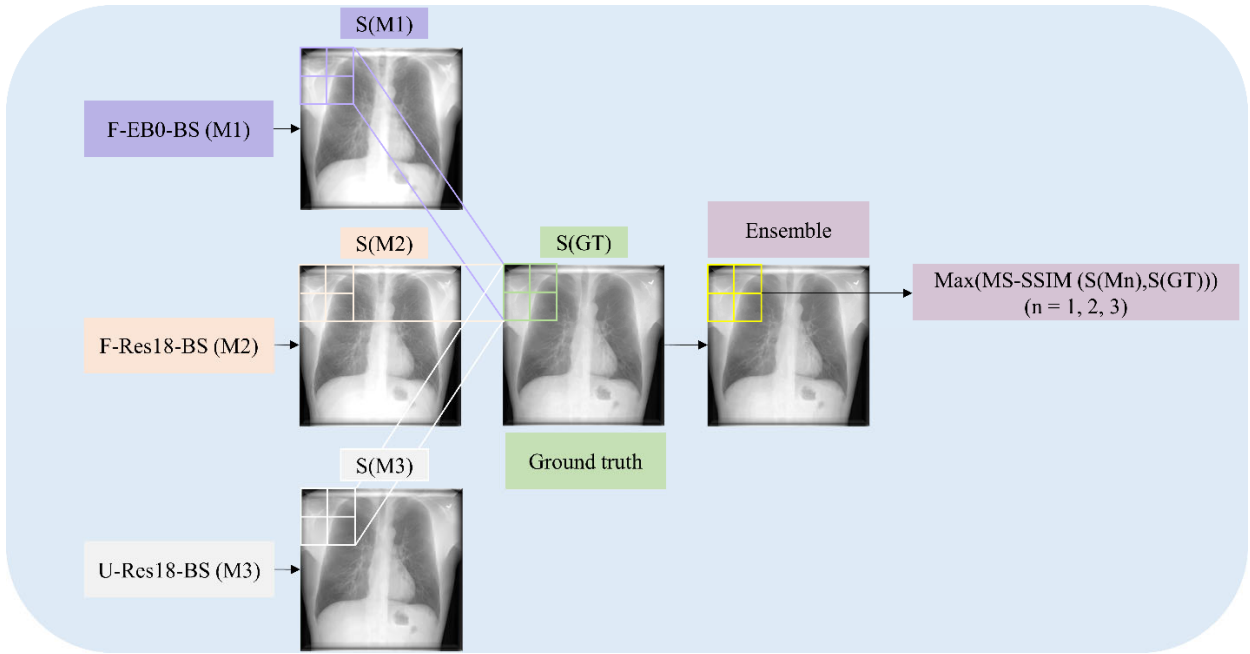

**Fig. 5.** The architecture of the proposed bone suppression model ensemble.

**Bone suppression model evaluation**

We performed evaluations by using the histograms of the ground truths and the bone-suppressed images predicted by the individual bone-suppression models and the ensemble. Several metrics such as correlation, intersection, chi-square distance, and Bhattacharyya distance are measured to investigate for similarity. The higher the value of correlation and intersection, the closer (or more similar) are the histograms of the image pairs. For distance-based metrics such as chi-square and Bhattacharyya, a smaller value indicates a superior match between the histogram pairs. This implies the histograms of the predicted bone-suppressed images closely match their respective ground truths. The mathematical formulations of these metrics can be found in the literature (Open Source Computer Vision, 2020). The average values of the aforementioned metrics are computed for each model and the ensemble and compared for statistical significance.

**Classification model**

For classification, we initially used a custom U-Net model proposed in (Rajaraman et al., 2020) to segment the lung region of interest (ROI) from the CXRs showing no abnormalities and other COVID-19-like manifestations. This approach ensures that the models learn relevant features from the lung ROI and not the surrounding context. The U-Net model is trained to generate 256×256 lung masks that are overlaid on the input CXRs to delineate the lung boundaries and cropped to a bounding box containing the lung pixels. The lung-cropped CXRs are further preprocessed to enhance image contrast by saturating the top 1% and bottom 1% of all pixel values. We further performed pixel normalization, centering, and standardization to reduce computational complexity during model training.

The encoder of the best-performing bone-suppression model is truncated and appended with the following layers: (i) Zero padding (ZP); (ii) Convolutional layer with 512 filters, each of size 3×3; (iii) Global average pooling (GAP), and (iv) a dense layer with two neurons and Softmax activation to classify the CXRs as showing normal lungs or COVID-19-related manifestations. This approach is followed to transfer the CXR modality-specific knowledge learned from the bone suppression task to improve performance in a relevant classification task. A study of the literature reveals several works that used CXR modality-specific models to transfer knowledge and improve classification and disease ROI localization performance in a relevant task (Islam et al., 2017; Rajaraman et al., 2020, 2019).

Recall that we use a collection of 3016 CXRs showing COVID-related manifestations and 3016 CXRs showing no abnormalities for the classification task. The best-performing bone suppression model is used to suppress the bones in this CXR collection. Since the ground truth soft-tissue images are not available for these CXRs, the bone suppression ensemble could not be used. We used 90% of this data for training and 10% for hold-out testing. We allocated 10% of the training data for validation with a fixed seed.

The classification model is then retrained individually on the non-bone-suppressed and bone-suppressed CXR images to classify them as showing no abnormalities or COVID-19-like manifestations. We performed augmentation with random affine transformations such as rotations (-10 to 10 degrees), horizontal and vertical pixel shifting (-5 to 5 pixels], zooming, and horizontal mirroring, to introduce variability into the training process and reduce model overfitting to the training data. The model is compiled using a stochastic gradient descent optimizer with an initial learning rate of 1e-3. We reduced the learning rate when the validation performance did not improve. We used callbacks to store model weights and early stopping after patience of 10 epochs to prevent overfitting and stored the best weights for further analysis. The best model is used to predict the test set and output class probabilities. The following metrics are measured to compare model performance: (i) accuracy; (ii) AUROC; iii) precision (P); (iv) recall (R); (v) AUPRC; (vi) F-score; and (vii) MCC. These metrics are expressed below.

$$Accuracy = \frac{TP + TN}{TP + TN + FP + FN}$$

$$Recall = \frac{TP}{TP + FN}$$

$$Precision = \frac{TP}{TP + FP}$$

$$F - score = 2 \times \frac{Precision \times Recall}{Precision + Recall}$$

$$MCC = \frac{TP \times TN - FP \times FN}{((TP + FP)(TP + FN)(TN + FP)(TN + FN))^{1/2}}$$

Here, TP, TN, FP, and FN denote the true positive, true negative, false positive, and false negative values, respectively. Additionally, we used our in-house class-selective relevance map (CRM) localization algorithm (Incheol Kim et al., 2019) to interpret the predictions of the model trained on non-bone-suppressed and bone-suppressed images and ensure they learned to highlight the disease ROI manifestations.

**Statistical analysis**

Statistical analysis is performed to investigate for a significant performance difference between the models. For bone suppression, we performed a one-way Analysis of Variance

(ANOVA) to analyze if a significant difference existed in the MS-SSIM and chi-square distance values obtained using the top-3 performing bone-suppression models and the ensemble. We performed Shapiro-Wilk and Levene tests to analyze if the prerequisite conditions of data normality and homogeneity of variances are satisfied to perform one-way ANOVA analyses. For classification, we measured the 95% binomial confidence intervals (CI) as the Exact Clopper-Pearson interval for the MCC metric to compare the classification performance achieved by the models trained on non-bone-suppressed and bone-suppressed images. We used R statistical software (Version 4.1.1) to perform these evaluations.

## III. Results

### Bone Suppression

Recall that the proposed bone suppression models are trained on the augmented NIH-CC-DES-Set 2 dataset and tested using the NIH-CC-DES-Set 1 collection (n = 27). The performance achieved by the bone suppression models is shown in Table 2.

**Table 2. Performance achieved by the proposed bone suppression models using the NIH-CC-DES-Set 1 test set.** The best performances are denoted by bold numerical values in the corresponding columns.

| Model | PSNR | SSIM | MS-SSIM | Correlation | Intersection | Chi-square | Bhattacharya |
|---|---|---|---|---|---|---|---|
| Autoencoder-BS | 33.1861± 3.5922 | 0.9371± 0.0310 | 0.9798± 0.0093 | 0.5949± 0.1800 | 8.4827± 1.4190 | 1.4279± 0.9773 | 0.4009± 0.0878 |
| ResNet-BS | 30.9168± 3.1286 | 0.9420± 0.0261 | 0.9817± 0.0092 | 0.5142± 0.1831 | 8.2680± 1.5036 | 2.6780± 1.6202 | 0.4281± 0.0884 |
| U-EB0-BS | 35.9098± 1.5674 | 0.9359± 0.0306 | 0.9795± 0.0084 | 0.6529± 0.1576 | 8.8000± 1.3606 | 0.9004± 0.6436 | 0.3813± 0.0845 |
| U-Res18-BS | 35.7993± 1.4498 | 0.9402± 0.0283 | 0.9809± 0.0080 | 0.6518± 0.1403 | 8.8879± 1.4312 | 0.9767± 0.4622 | 0.3796± 0.0833 |
| U-SE-Res18-BS | 35.531± 1.6773 | 0.9325± 0.0310 | 0.9773± 0.0077 | 0.6421± 0.1505 | 8.6794± 1.3098 | 1.0484± 0.8215 | 0.383± 0.0836 |
| U-D121-BS | 33.7751± 1.3033 | 0.9284± 0.0301 | 0.9746± 0.0083 | 0.6017± 0.1543 | 8.4233± 1.6595 | 1.7434± 1.2997 | 0.3852± 0.0838 |
| U-IV3-BS | 34.8914± 1.7280 | 0.9368± 0.0294 | 0.9795± 0.0089 | 0.6411± 0.1339 | 8.8026± 1.4659 | 1.1195± 0.4987 | 0.3836± 0.0816 |
| U-MobileV2-BS | 27.6842± 0.1715 | 0.8593± 0.0342 | 0.9136± 0.0139 | 0.2583± 0.1131 | 5.7133± 1.5060 | 10.9967± 4.2341 | 0.4704± 0.0631 |
| FPN-EB0-BS | **36.5525±** | **0.9449±** | **0.9840±** | **0.6654±** | **9.0462±** | **0.6893±** | **0.3790±** |
| | 1.6923 | 0.0290 | 0.0081 | 0.1473 | 1.4529 | 0.4005 | 0.0846 |
| FPN-Res18-BS | 36.3233± 1.7004 | 0.9428± 0.0281 | 0.9823± 0.0079 | 0.6417± 0.1424 | 8.8840± 1.4194 | 0.9392± 0.3799 | 0.3856± 0.0833 |
| FPN-SE-Res18-BS | 36.0318± 1.6900 | 0.9418± 0.0294 | 0.9821± 0.0084 | 0.6334± 0.1559 | 8.8531± 1.4131 | 1.0227± 0.5185 | 0.3853± 0.0841 |
| FPN-D121-BS | 35.2788± 1.4938 | 0.9402± 0.0283 | 0.9794± 0.0082 | 0.6290± 0.1365 | 8.7087± 1.5015 | 1.2203± 0.9092 | 0.3827± 0.0818 |
| FPN-IV3-BS | 33.7446± 1.8066 | 0.9369± 0.0310 | 0.9793± 0.0084 | 0.6225± 0.1560 | 8.6645± 1.3670 | 1.1846± 0.7676 | 0.3910± 0.0817 |
| FPN-MobileV2-BS | 33.5028± 1.3452 | 0.9255± 0.0320 | 0.9734± 0.0088 | 0.5767± 0.1743 | 8.1361± 1.6643 | 2.3053± 1.4224 | 0.3877± 0.0844 |

It is observed from Table 2 that the FPN model with the EfficientNet-B0 encoder backbone (FPN-EB0-BS) demonstrated superior performance for all metrics compared to other models. It is observed from Fig. 6 that all models predicted bone suppressed images that demonstrated substantial suppression of the bony structures. we performed quantitative analysis to differentiate these performances. In this regard, we observed that the FPN-EB0-BS model demonstrated the least values for the chi-square and Bhattacharya distances and superior values for the correlation and intersection measures. Higher values for the correlation and intersection metrics demonstrate that the bone-suppressed images predicted by the F-EB0-BS model closely match that of the ground truth soft-tissue images. Considering the chi-square and Bhattacharyya distance-based metrics, a smaller value indicates a superior match between the images. This infers that the F-EB0-BS model delivered superior bone suppression performance such that the predicted bone-suppressed image is highly similar to the ground truth. This performance is followed by the FPN model with ResNet-18 encoder backbone (F-Res18-BS) and the U-Net model with the ResNet-18 encoder backbone (U-Res18-BS) that demonstrated markedly improved values for the PSNR, SSIM, MS-SSIM, correlation, intersection, chi-square, and Bhattacharya distance measures compared to other models. These top-3 performing models are further considered to construct the ensemble. Fig. 6 shows the bone-suppressed images predicted using the proposed bone suppression models for an input CXR instance from the test set.

The predicted bone-suppressed images by the top-3 performing models are divided into sub-blocks of M×M dimensions. We empirically determined the value of M [4, 8, 16, 32, 64, 128, 256] that deliver superior bone suppression performance. For a given sub-block size, and in each sub-block, (i) we measured the MS-SSIM score between the sub-block of the bone-suppressed image predicted by each of the top-3 performing models and the corresponding sub-block of its respective ground truth, and (ii) performed a majority voting of the MS-SSIM score for each sub-block to identify the sub-block with the maximum MS-SSIM score and use it in constructing the final bone-suppressed image. Table 3 shows the performance achieved while constructing the ensemble model using varying sub-block sizes.

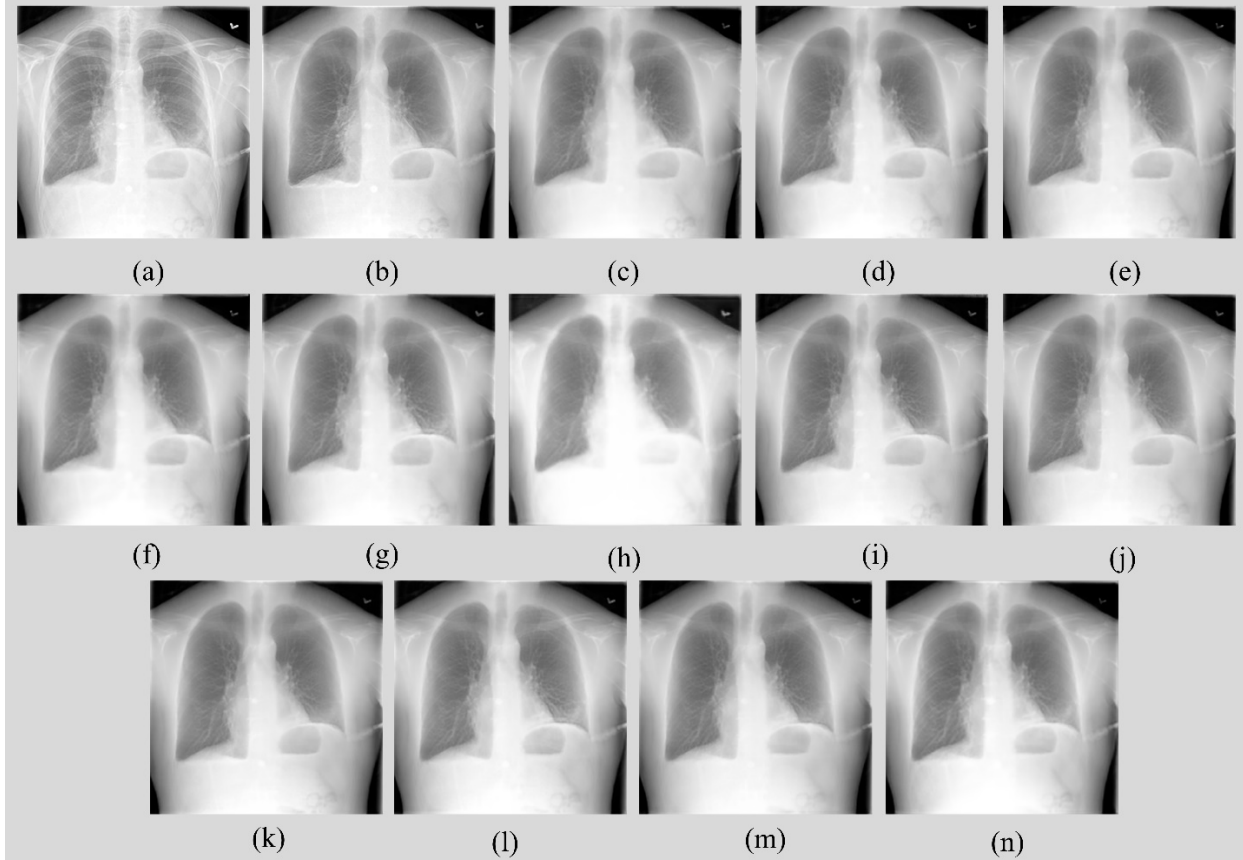

**Fig. 6. Bone-suppressed CXR images predicted by the proposed models using a CXR sample from the NIH–CC DES-Set 1 test set.** (a) Original CXR; (b) Ground truth soft tissue image; (c) U-EB0-BS; (d) U-Res18-BS; (e) U-SE-Res18-BS; (f) U-D121-BS; (g) U-IV3-BS; (h) U-MobileV2-BS; (i) FPN-EB0-BS; (j) FPN-Res18-BS; (k) FPN-SE-Res18-BS; (l) FPN-D121-BS; (m) FPN-IV3-BS and (n) FPN-MobileV2-BS.

**Table 3. Performance achieved by the bone suppression model ensemble using various sizes for the sub-blocks.**
The best performances are denoted by bold numerical values in the corresponding columns.

| Block size | PSNR | SSIM | MS-SSIM | Correlation | Intersection | Chi-Square | Bhattacharya |
|---|---|---|---|---|---|---|---|
| 4×4 | **36.7977± 1.6207** | **0.9465± 0.0272** | **0.9848± 0.0073** | **0.6720± 0.1404** | **9.0862± 1.4413** | **0.6174± 0.2726** | **0.3778± 0.0839** |
| 8×8 | 36.4574± | 0.9226± | 0.8721± | 0.6344± | 8.5230± | 1.2636± | 0.3806± |
| | 1.4724 | 0.0255 | 0.0223 | 0.1361 | 1.3419 | 0.4771 | 0.0822 |
| 16×16 | 36.7651± 1.6012 | 0.9437± 0.0256 | 0.9837± 0.0073 | 0.6598± 0.1431 | 9.0193± 1.4464 | 0.7282± 0.3169 | 0.3800± 0.0839 |
| 32×32 | 35.7137± 1.2588 | 0.8965± 0.0266 | 0.8390± 0.0237 | 0.5161± 0.1218 | 7.1297± 1.1228 | 3.4226± 1.0079 | 0.3901± 0.0801 |
| 64×64 | 36.2657± 1.4698 | 0.9218± 0.0272 | 0.8719± 0.0221 | 0.6297± 0.1409 | 8.4949± 1.3528 | 1.3402± 0.5874 | 0.3815± 0.0834 |
| 128×128 | 36.4872± 1.5982 | 0.9380± 0.0282 | 0.9213± 0.0174 | 0.6667± 0.1483 | 8.9424± 1.4365 | 0.7807± 0.4931 | 0.3784± 0.0850 |
| 256×256 | 36.5787± 1.6885 | 0.9458± 0.0284 | 0.9841± 0.0080 | 0.6641± 0.1470 | 9.0304± 1.4599 | 0.7026± 0.4129 | 0.3796± 0.0849 |

It is observed from Table 3 that the ensemble performance with various sub-block sizes is superior compared to the performance achieved using the top-3 performing models (from Table 2). We observed that using a sub-block size of 4×4, the ensemble of the top-3 performing models achieved superior performance in terms of PSNR, SSIM, MS-SSIM, correlation, intersection, chi-square, and Bhattacharya distances compared to using other sub-block sizes and the top-3 performing models. This performance excellence could be due to the increased granularity, i.e., the level of comparing the pixel details while measuring the MS-SSIM score in small sub-blocks of 4×4 dimensions.

We performed a one-way ANOVA analysis to observe if a statistically significant difference existed in the MS-SSIM and chi-square values obtained using the ensemble with sub-block size 4×4, and the top-3 performing bone-suppression models namely, the F-EB0-BS, F-Res18-BS, and U-Res18-BS models. Fig. 7 shows the mean plots for the MS-SSIM and chi-square values, respectively, obtained by the models. The one-way ANOVA analyses require that the assumptions regarding the normal distribution of the data and homogeneity of data variances are satisfied. We performed the Shapiro-Wilk normality test and Levene test for analyzing the homogeneity of variances. For the MS-SSIM metric, we observed that the *p*-values for the Levene ($p=0.9828$) and Shapiro-Wilk ($p=0.3824$) tests are not statistically significant ($p>0.05$). This confirms that the assumptions of data normality and homogeneous variances are satisfied. Hence, we performed one-way ANOVA analyses. One-way ANOVA is performed by measuring the size of the group, the variance within groups, and the variance between the means of the groups. This information is collectively used to measure the F statistic. In this study, we have four groups/models (i.e., the 4×4 ensemble, F-EB0-BS, F-Res18-BS, and U-Res18-BS models) with

27 observations each, hence the distribution is mentioned as F(3, 104). Considering the MS-SSIM metric, we observed that no statistically significant difference existed between the 4×4 ensemble and the top-3 performing models (F(3, 104)=0.886, *p*=0.451, *p*>0.05).

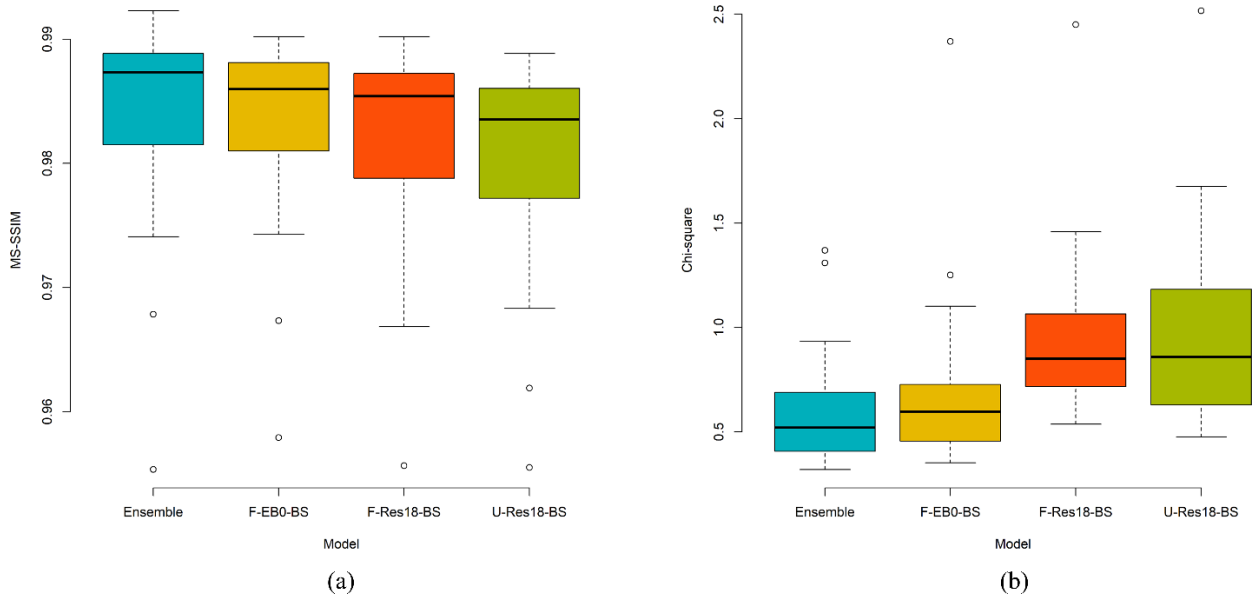

(a)                                                                                          (b)

**Fig. 7. Statistical analyses using one-way ANOVA.** (a) and (b) shows the mean plot for the MS-SSIM and chi-square values, respectively, obtained by the ensemble (4×4), FPN_EB0, FPN_Res18, and U-Res18 models.

A similar analysis is performed using the chi-square distance metric. We observed that the conditions of data normality and homogeneous variances are satisfied based on the *p*-values obtained using the Shapiro-Wilk (*p*=0.4768) and Levene (*p*=0.4321) tests (*p*>0.05). The one-way ANOVA analysis revealed that a statistically significant difference existed in the chi-square values obtained using the 4×4 ensemble, F-EB0-BS, F-Res18-BS, and U-Res18-BS models (F(3, 104)=5.838, *p*=0.001, *p*<0.05). We further performed Tukey post-hoc analyses to identify the models that demonstrate these significant differences in the chi-square values. We observed that the chi-square distance value obtained using the 4×4 ensemble (0.6174±0.2726) is significantly smaller compared to the F-Res18-BS (0.9392±0.3799, p = 0.0142) and U-Res18-BS (0.9767±0.4622, p=0.0047) models. Also, the chi-square value obtained using the F-EB0-BS model is significantly smaller (*p*=0.0355) compared to the U-Res18-BS model. These evaluations underscored the fact that the 4×4 ensemble achieved significantly smaller values for the chi-square metrics (*p*<0.05). Unlike the top-3 performing models, the bone-suppressed images predicted by the ensemble closely resembled the ground truth soft-tissue images. Recall that the best-performing F-EB0-BS bone suppression model is used to suppress the bones in the CXRs used in this classification task. This is because the ground truth soft-tissue images are not available for these CXRs. Hence, the bone suppression ensemble could not be used. Fig. 8 shows the bone-suppressed images predicted by the F-EB0-BS model for instances of CXRs showing COVID-19-related manifestations.

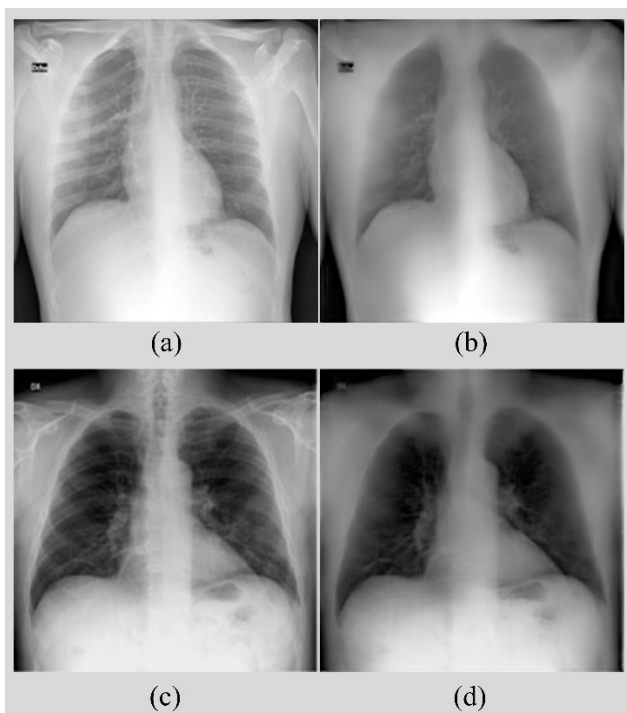

**Fig. 8. Bone-suppressed images predicted by the F-EB0-BS model using instances of CXRs showing COVID-19-related manifestations.** (a) CXR from the BIMCV-COVID19+ CXR data collection; (b) Corresponding bone-suppressed image; (c) CXR from the Twitter COVID-19 collection, and (d) Corresponding bone-suppressed image.

It is observed that the F-EB0-BS model generalized to the unseen CXRs from the classification data that are not used during bone-suppression model training and validation. It is observed that the bony structures are well suppressed while preserving the image resolution.

## Classification

Recall that the encoder of the best-performing F-EB0-BS bone suppression model is truncated and added with the classification layers to classify the CXRs as showing normal lungs or COVID-19-like manifestations. Such an approach is followed to transfer CXR modality-specific knowledge to improve classification performance. The classification model is retrained on the non-bone-suppressed and bone-suppressed CXR images, and the measured performance is shown in Table 4, and illustrated in Fig. 9 in terms of AUROC, confusion matrix, normalized Sankey diagram, and AUPRC curves.

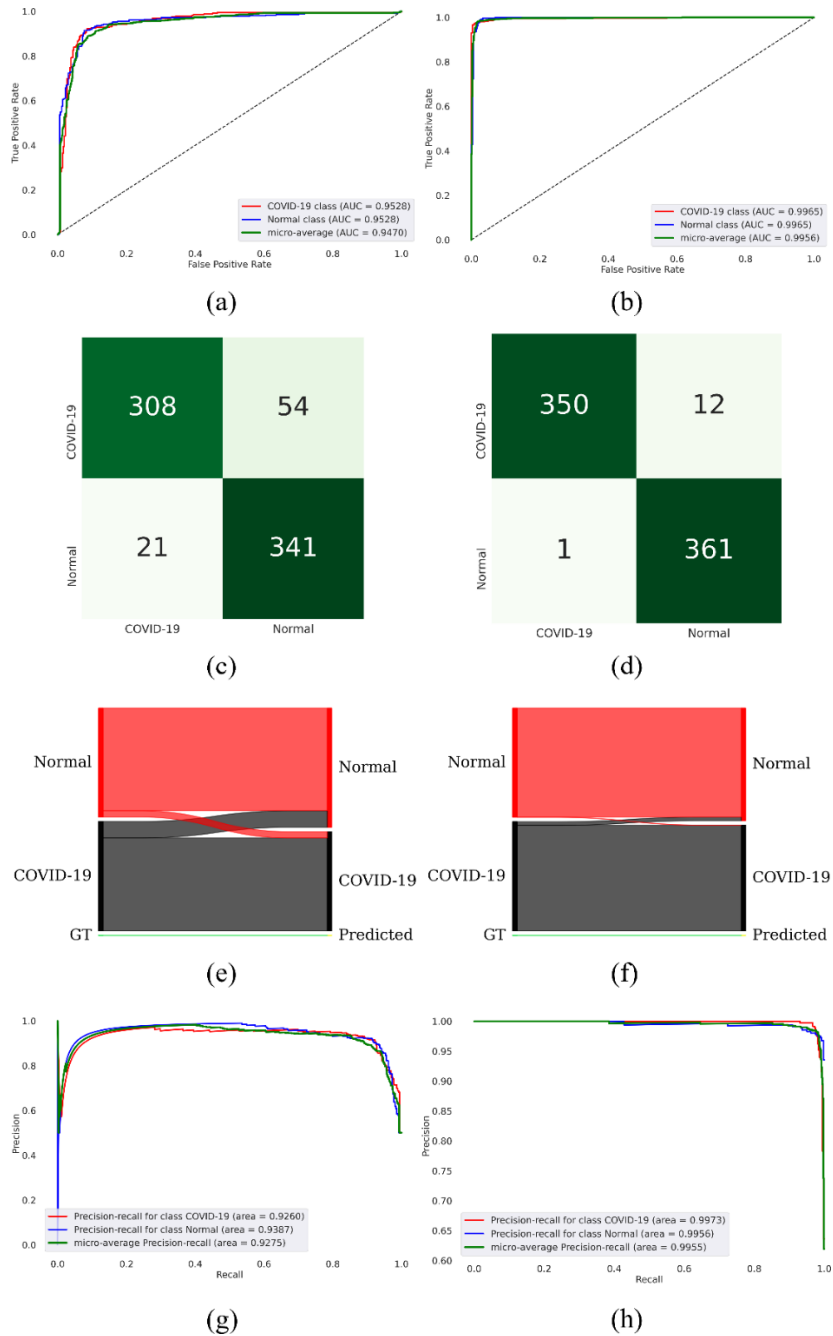

**Fig. 9. Classification performance achieved by the model trained on non-bone-suppressed and bone-suppressed images.** (a), (c), (e), and (g) denote the AUROC, confusion matrix, Sankey diagram, and AUPRC curves achieved

through training the model with non-bone-suppressed images; (b), (d), (f), and (h) denote the AUROC, confusion matrix, Sankey diagram, and AUPRC curves achieved through training with the bone-suppressed images.

**Table 4. Classification performance achieved with the model trained on non-bone-suppressed and bone-suppressed images.** Data in parenthesis denote the 95% binomial CI measured as the Exact Clopper Pearson interval for the MCC metric. Bold numerical values denote superior performance in respective columns.

| Data | Accuracy | AUROC | AUPRC | Sensitivity | Precision | F-score | MCC |
|---|---|---|---|---|---|---|---|
| Non bone suppressed | 0.8964 | 0.9470 | 0.9275 | 0.8964 | 0.8997 | 0.8962 | 0.7961 (0.7667, 0.8255) |
| Bone suppressed | **0.9820** | **0.9980** | **0.9981** | **0.9820** | **0.9825** | **0.9820** | **0.9645** **(0.9510, 0.9780)** |

We observed from Table 4 and Fig. 9 that the classification model trained on bone-suppressed images demonstrated superior performance in terms of accuracy, AUROC, AUPRC, sensitivity, precision, F-score, and MCC metrics, compared to the model trained on non-bone-suppressed images. The 95% binomial CI value obtained for the MCC metric using the model trained on bone-suppressed images demonstrated a tighter error margin, higher precision, and is found to be significantly superior ($p < 0.05$) compared to the MCC metric achieved by the model trained on non-bone-suppressed images.

We qualitatively evaluated the performance of the models trained on non-bone-suppressed and bone-suppressed images to ensure if the models learned to highlight COVID-19 disease-specific ROIs and not the surrounding context. We used the CRM localization tool to interpret model behavior. Fig. 10 shows the instances of CXRs, and the CRM-based disease ROI localization obtained using the trained models. Fig. 10a, Fig. 10d, and Fig. 10g show instances of CXRs from the Twitter COVID-19 CXR collection with expert annotations shown in blue bounding boxes. Fig. 10b, Fig. 10e, and Fig. 10h show the localization achieved using the model trained on non-bone-suppressed images. It could be observed that the model is highlighting the surrounding context but not COVID-19-specific disease manifestations. This demonstrates that the model has not learned meaningful features regarding COVID-19 manifestations. Fig. 10c, Fig. 10f, and Fig. 10i show the localization achieved using the model trained on bone-suppressed images. We could observe that this model precisely highlighted regions specific to COVID-19 manifestations, thereby demonstrating that the model learned meaningful features, confirming the experts' knowledge about the disease.

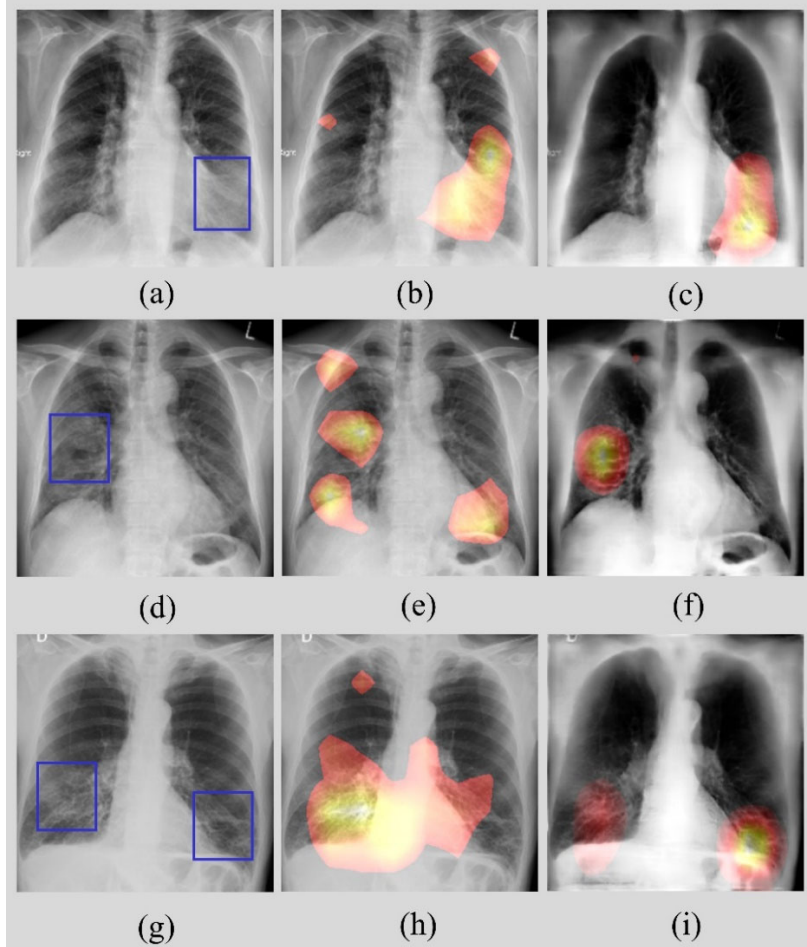

**Fig. 10. CRM-based localization of COVID-19-related manifestations.** (a) (d) and (g) denote instances of CXRs from the Twitter COVID-19 CXR collection showing COVID-19-manifestations with expert annotations (shown with blue bounding boxes); (b) (e) and (f) shows the regions highlighted by the model trained on non-bone-suppressed images; (c) (f) and (i) shows the COVID-19-consistent ROIs highlighted by the model trained on bone-suppressed images.

## IV. Discussion

The observations made from this retrospective study underscores the need for (i) customizing a model for the problem under study, (ii) constructing a model ensemble for bone suppression, and (iii) interpreting model behavior.

Our proposed approach facilitates predicting a bone-suppression image given an input CXR image. This is computationally effective than other studies proposed in the literature (Kodama et al., 2018; Li et al., 2011b; Matsubara et al., 2019; Suzuki et al., 2006; Yang et al., 2017) that propose a series of steps to generate bone-selective images and subtract them from input

CXRs to increase soft-tissue visibility. A limitation of this approach is that a sub-optimal generation of bone-selective images introduces noise and distortion into the process and may adversely impact decision-making. We proposed several custom models and experimented with the SOTA architectures like U-Nets and FPN using various ImageNet-pretrained encoder backbones to obtain superior bone suppression performance. To the best of our knowledge, this study is the first to explore the use of these models in the context of an image denoising problem where the bony structures are considered noise in an input CXR. Through extensive empirical evaluations, we observed that the FPN model with the EfficientNet-B0 encoder backbone delivered superior bone suppression performance, followed by the FPN model with ResNet-18, and U-Net with ResNet-18 encoder backbones. The bone-suppressed images predicted by these top-3 models appeared sharp while preserving soft-tissue characteristics. Therefore, these images could be used for further CXR image analysis such as screening for cardiopulmonary diseases. We propose an ensemble approach toward bone suppression that demonstrated superior values for PSNR, SSIM, MS-SSIM, correlation, intersection, chi-square, and Bhattacharya distance metrics compared to other models. This underscores the fact the ensemble model improved bone suppression performance so that the predicted bone suppressed image closely matched the ground-truth, soft-tissue image.

We observed the effect of bone suppression toward improving COVID-19 detection using CXRs. We observed that the classification model trained using bone-suppressed images demonstrated significantly superior performance in terms of accuracy, AUPRC, AUROC, precision, recall, F-score, and MCC, compared to the model trained on non-bone-suppressed images. We further observed through localization studies that the models trained on bone-suppression images precisely highlighted COVID-19-related ROI in the input CXR, confirming the expert knowledge of the disease. This underscores the fact that, unlike the model trained on non-bone-suppressed images, the models trained on bone-suppressed images learned meaningful features and not the surrounding context, to classify the CXRs to their respective classes. The models trained on non-bone-suppressed images are accurate, however, they demonstrated sub-optimal COVID-19-consistent ROI localization. This underscores the fact that the disease-specific ROI location ability of a trained model is not related to its classification accuracy and localization studies are therefore indispensable to interpret model behavior.

This study suffers from the following limitations: (i) We used the best-performing bone-suppression model, and not the model ensemble, to suppress bones in the CXR data used for the classification task. This is because we do not have the ground truth soft-tissue images for the CXRs. We could benefit from the bone suppression ensemble if the CXRs are accompanied by their soft-tissue counterparts. The lack of large-scale publicly available bone-suppressed CXR datasets is a limitation in training the bone-suppression models. The studies reported in the literature (Kodama et al., 2018; Li et al., 2011b; Matsubara et al., 2019; Suzuki et al., 2006; Yang et al., 2017) used JSRT CXR images and their bone-suppressed counterparts generated by an automated algorithm developed by the researchers from the Budapest University of Technology and Economics (Budapest University of Technology and Economics (BME), 2013) to train the bone suppression models. However, automated algorithms could introduce noise and artifacts into the bone suppression process, thereby leading to sub-optimal model training and inference. To the best of our knowledge, this is the first study to use DES CXRs to train the bone-suppression models. However, this is not large-scale data and hence may not encompass a wide range of variability in the bone structures. It would be possible to propose deeper architectures with increased availability of DES CXRs that would introduce sufficient data diversity into the training process and improve model confidence and performance generalization to real-world data. This is not a classification-related study, but we aim to demonstrate if bone suppression would improve performance toward COVID-19 detection. Choosing the best classification model is beyond the scope of this study. The proposed bone-suppression approach could be extended to other image denoising problems. The importance of using bone suppressed CXRs for detecting other cardiopulmonary abnormalities including lung nodules, TB, pneumonia, among others would be a good research avenue. We believe our results will improve human visual interpretation of COVID-19 related findings, as well as automated detection in AI-driven workflows.